\documentclass{article}
\usepackage{etoolbox}
\patchcmd{\thebibliography}{\section*{\refname}}{}{}{}
\patchcmd{\thebibliography}{\addcontentsline{toc}{section}{\refname}}{}{}{}
\usepackage{authblk}

\usepackage{spconf,amsmath,graphicx}
\usepackage{gensymb}
\usepackage[ruled,linesnumbered]{algorithm2e}
\usepackage{subfig}
\usepackage{mathtools}

\DeclarePairedDelimiter\floor{\lfloor}{\rfloor}

\usepackage{collcell}
\usepackage{xstring}
\usepackage{datatool}
\usepackage{booktabs}
\usepackage{environ}

\newcommand{\Small}{\fontsize{11.5pt}{20pt}\selectfont}

\title{Prediction of treatment outcome for autism from structure of the brain based on sure independence screening}
\name{
\Small{
Juntang Zhuang$^{1}$, Nicha C. Dvornek$^{2,3}$, Qingyu Zhao$^{4}$, Xiaoxiao Li$^{1}$, Pamela Ventola$^{2}$, James S. Duncan$^{1,3}$
}
}
\Small{
\address{
$^1$ Biomedical Engineering, Yale University, New Haven, CT, USA \\
$^2$ Child Study Center, Yale University, New Haven, CT, USA \\
$^3$ Radiology and Biomedical Imaging, Yale School of Medicine, New Haven, CT, USA \\
$^4$ Department of Psychiatry and Behavioral Sciences, Stanford University,Stanford, USA
}
}
\begin{document}
%
\maketitle
\begin{abstract}
Autism spectrum disorder (ASD) is a complex neurodevelopmental disorder, and behavioral treatment interventions have shown promise for young children with ASD. However, there is limited progress in understanding the effect of each type of treatment. In this project, we aim to detect structural changes in the brain after treatment and select structural features associated with treatment outcomes. The difficulty in building large databases of patients who have received specific treatments and the high dimensionality of medical image analysis problems are the challenges in this work. To select predictive features and build accurate models, we use the sure independence screening (SIS) method. SIS is a theoretically and empirically validated method for ultra-high dimensional general linear models, and it achieves both predictive accuracy and correct feature selection by iterative feature selection. Compared with step-wise feature selection methods, SIS removes multiple features in each iteration and is computationally efficient. Compared with other linear models such as elastic-net regression, support vector regression (SVR) and partial least squares regression (PSLR), SIS achieves higher accuracy. We validated the superior performance of SIS in various experiments: First, we extract brain structural features from FreeSurfer, including cortical thickness, surface area, mean curvature and cortical volume. Next, we predict different measures of treatment outcomes based on structural features. We show that  SIS achieves the highest correlation between prediction and measurements in all tasks. Furthermore, we report regions selected by SIS as biomarkers for ASD.
\end{abstract}
\begin{keywords}
Sure independence screening, autism, brain structure
\end{keywords}
\vspace{-0.2cm}
\section{Introduction}
\label{sec:intro}
Autism spectrum disorder (ASD) is a neurodevelopmental disorder characterized by deficits in social communications and repetitive behaviors \cite{lord2000autism}. Structural changes of the brain have been identified in patients with ASD in the literature:  Sparks et al. found children with ASD have significantly increased cerebral volumes compared with typically developing (TD) and delayed developing (DD) children \cite{sparks2002brain}. Pereira et al. found changes in cingulate gyrus, paracentral lobule and superior frontal gyrus on subjects with ASD \cite{pereira2018differences}.
\par
Behavioral-based treatments are widely used for ASD, and Pivotal Response Treatment (PRT) is an empirically supported therapy \cite{koegel2006pivotal} which addresses core deficits in social communication skills. Although structural changes in the brain have been studied for ASD, there has been limited progress in identifying the effect of PRT on brain structure. Therefore, predictive models for treatment outcomes based on brain structural changes is essential for understanding the mechanism of ASD. Furthermore, accurate predictive models are more robust than analytical models which tend to overfit to small datasets.
\par
The difficulty in building large databases and the high dimensionality of medical images are the main challenges. We introduce sure independence screening (SIS) \cite{fan2008sure}, a feature selection method for ultra-high dimensional general linear models. Although the screening method is widely used in genetics research, the neuroscience community tends to use simpler linear models such as elastic-net \cite{zou2005regularization}. However, these simple models cannot deal with ultra-high dimensional problems, yet more straightforward step-wise feature selection methods are computationally expensive. In this paper, we demonstrate the superior performance of SIS over other models in the prediction of changes in severity score based on structural features of the brain. Furthermore, we analyze selected features as biomarkers for ASD.   
\vspace{-0.2cm}
\section{Methods}
\vspace{-0.2cm}
\subsection{Difficulties for high-dimensional problems}
\label{difficulties}
The high dimensional problem refers to problems where the dimension $p$ is larger than the sample size $n$. The high dimensionality causes the following problems: (a) Design matrix $X$ has more columns than rows, which causes matrix $X^TX$ to be singular and large. Linear regression with no constraint will generate an infinite number of solutions to training data, and it's hard to determine which is the correct model and generalizes to test data. (b) In a high dimensional case, an unimportant variable can have a high correlation with the response variable or predictive variables, and this adds to the difficulty of variable selection. (c) The high dimension $p$ makes step-wise feature selection methods computationally infeasible.
\par
There are mainly three types of feature selection methods: wrapper methods, embedded methods, and filter methods. Wrapper methods such as forward selection, backward elimination, and recursive feature elimination have a huge computational burden because each subset of features has to be tested. Embedded methods with built-in feature selection, such as LASSO and elastic-net regression \cite{zou2005regularization}, usually cannot deal with ultra-high dimensional problems. Filter methods usually pre-select variables based on some importance measures before learning; however, this pre-processing step usually cannot accurately select predictive features. Zhuang et al. proposed a two-level feature selection approach specially for brain image analysis \cite{zhuang2018prediction2}; however, the region-level feature selection in their method depends on the assumption of local smoothness of the input image, therefore is suited for voxel-wise features but not ROI-wise features. Zhuang et al. proposed a non-linear feature selection method based on random forest \cite{zhuang2018prediction1}; however, their method is well suited for non-linear models but not linear models. Therefore, we introduce SIS \cite{fan2008sure}, a fast and accurate feature selection model for ultra-high dimensional general linear models.

\vspace{-0.3cm}
\subsection{Sure independence screening}
\label{SIS}
Sure screening means a property that all the important variables survive after applying a variable screening procedure with probability tending to 1 (Theorem 3 in \cite{fan2008sure}) as sample size $n$  increases. This asymptotic property gives a theoretical guarantee on the performance of SIS.
\par
\begin{figure}[!htb]
    \centering
    \includegraphics[width=0.8\linewidth]{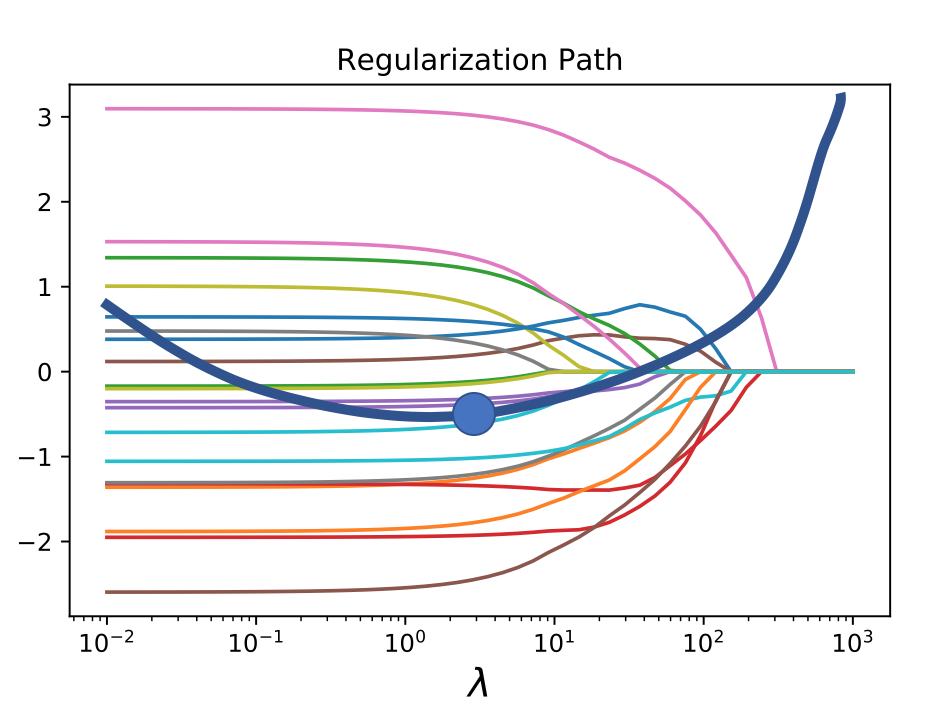}
    \caption{X axis shows different values of regularization strength $\lambda$. Colored lines show the regularization path, and the thick blue line demonstrates the loss for some model selection criterion (e.g. AIC). The blue dot represents the $\lambda$ with the lowest loss.}
    \label{regularization_path}
\end{figure}

Suppose the design matrix $X$ is centered and normalized to unit variance for each column, and the dimension of $X$ is $n \times p$, where $n$ is the number of observations, and $p$ is the number of variables. Denote response variable as $y$, and $y$ is a vector of length $n$. Denote the true predictive variables as $M_\ast = \{ 1 \leq i \leq p: \beta_i \neq 0\}$, where the true model is $Y=X\beta$. Denote the non-sparsity rate as $s=\vert M_\ast \vert$. A component-wise regression is defined as: 
\begin{equation}
\label{compoentreg}
w = X^T y
\end{equation}
Note that each column of $X$ is already normalized. The $i$th component of $w$, denoted as $w_i$, is proportional to the linear correlation between $X_i$ and $y$. For a given parameter $\gamma \in (0,1)$, the $\floor{\gamma n}$ largest components in $w$ are selected. The subset of preserved features are defined as:
\begin{equation}
\label{susbet}
M_\gamma = \{ 1 \leq i \leq q: \vert w_i \vert \text{ is among the largest} \floor{\gamma n}\}
\end{equation}
 The full algorithm is summarized in algorithm \ref{algo} and Fig. \ref{regularization_path}. The iterative SIS algorithm performs feature selection recursively, until some criterion is satisfied. Within each iteration, for a given penalty (e.g. $l_1$ penalty for LASSO, or Minimax Concave Penalty (MCP) ), by varying the regularization strength, there's a set of corresponding solutions summarized as ``regularization path". For each point along the path, the goodness of fitting can be measured by some criterion (e.g. Akaike information criterion) as shown in the thick blue curve in Fig. \ref{regularization_path}. The selected model is represented by the dot in Fig. \ref{regularization_path}, and the corresponding features are selected for the next iteration.
 \vspace{-0.2cm}
\begin{algorithm}[h]
\SetAlgoLined
 Set $\gamma \in (0,1)$\;
 $M^0_{keep} \gets M$\;
 $p_{keep} \gets p$\;
 $i \gets 0$\;
 \While{ $p_{keep}\geq n$ }{
       $i \gets i+1$\;
       Select model $M^i_{keep} \subset M^{i-1}_{keep}$:
       \newline
       \indent (1) Calculate the regularization path for a certain sparsity penalty (e.g. $l_1$ penalty ) varying with different regularization parameter $\lambda$ \newline
       \indent (2) Perform model selection along the regularization path according to some criterion (e.g. AIC) \;
       Update $p_{keep}$\;
 }
 \caption{Iterative sure independence screening algorithm}
 \label{algo}
\end{algorithm}
\vspace{-0.8cm}
\vspace{-0.6cm}
\subsection{Sparsity penalties for general linear models}
Within each iteration in Algorithm 1, the algorithm fits a generalized linear model. Different sparsity penalties can be applied considering the computation capability, and this problem can be efficiently solved with standard algorithms such as coordinate descent.
\par
We briefly introduce two types of sparsity penalty in this paragraph. $l1$ penalty is widely used for feature selection and reduces the problem to LASSO,  and its variant adaptive LASSO \cite{zou2006adaptive} introduces the sparsity penalty term as 
\begin{equation}
\lambda \sum_{j=1}^d a_j \vert \beta_j \vert
\end{equation}
where $a_j$ is the weight for the $j$th component, and $\lambda$ is a pre-defined penalty hyper-parameter.
\par 
Another commonly used penalty is Minimax concave penalty (MCP) \cite{zhang2010nearly}, where the problem is defined as:
\begin{equation}
\beta = \text{argmin}_{\theta} \big \{ \frac{1}{2n}\sum_{i=1}^n(Y_i-x_i^T\theta)^2 + \sum_{j=1}^d p_{\lambda_j} (\vert \theta_j \vert) \big \}
\end{equation}
where $p_{\lambda}(\vert \theta \vert)$ is defined as:
\begin{equation}
p_{\lambda}(\vert \theta \vert) = (a \lambda - \vert \theta \vert)_+ / a \text{,   for some } a>2
\end{equation}
For any penalties, the regularization path can be calculated, which is then used to select the optimal regularization strength.
\vspace{-0.3cm}
\subsection{Model selection criterion}
After calculating the regularization path, the optimal regularization strength can be selected by some model selection criterion. The loss for model selection is shown in thick blue curve in Fig. \ref{regularization_path}, and the optimal point is marked with a dot. The model selection criterion typically balances goodness of fitting training data and complexity of the model. We introduce three types of criterion here. 
\par
(a) Akaike information criterion (AIC) \cite{bozdogan1987model}, where the parameters are chosen as 
\begin{equation}
\label{aic}
\beta = \text{argmin}_{\theta} \big \{ - log P(y\vert X, \theta) + 2 d_f \big \}
\end{equation}
where $d_f$ is the degree of freedom of the fitted model, and $\theta$ is the parameter to be estimated.
\par
(b) Bayesian information criterion (BIC) \cite{chen1998speaker}, where the parameters are chosen as
\begin{equation}
\label{bic}
\beta = \text{argmin}_{\theta} \big \{ - log P(y\vert X, \theta) + log( n) d_f \big \}
\end{equation}
\par
(c) Extended bayesian information criterion (EBIC) \cite{chen2008extended}, where the model is determined as
\begin{equation}
\label{ebic}
\beta = \text{argmin}_{\theta} \big \{ - log P(y\vert X, \theta) + log( n) d_f + 2 \eta log( C_p^{d_f}) \big \}
\end{equation}
where $\eta$ is a pre-defined parameter, and $C_p^{d_f}$ represents number of choices to choose $d_f$ variables from a total of $p$ variables. 
\begin{table*}[ht]
\scalebox{0.735}{
\begin{tabular}{|l|l|l|l|l|l|l|l|l|l|l|l|l|l|l|l|l|}
\hline
Task        & \multicolumn{4}{l|}{SRS}                     & \multicolumn{4}{l|}{ADOS calibrated severity} & \multicolumn{4}{l|}{ADOS social affect}    & \multicolumn{4}{l|}{ADOS restricted and repetitive behavior} \\ \hline
Method      & \textbf{SIS}   & PLS   & Elastic-net & SVR   & \textbf{SIS}   & PLS    & Elatic-net  & SVR   & \textbf{SIS}  & PLS  & Elastic-net & SVR   & \textbf{SIS}     & PLS              & Elastic-net   & SVR    \\ \hline
RMSE        & \textbf{13.34} & 18.06 & 15.91       & 17.92 & \textbf{2.03}  & 2.10   & 2.23        & 2.73  & \textbf{3.18} & 3.85 & 3.74        & 4.48  & 2.28             & \textbf{2.12}    & 2.40          & 2.12   \\ \hline
Correlation & \textbf{0.71}  & -0.09 & 0.39        & -0.20 & \textbf{0.44}  & -0.15  & -0.36       & -0.64 & \textbf{0.67} & -0.2 & 0.22        & -0.78 & \textbf{0.51}    & 0.20             & 0.11          & 0.47   \\ \hline
\end{tabular}
}
\label{table1}
\vspace{-0.1cm}
\end{table*}

\vspace{-0.3cm}

\begin{table*}[]
\scalebox{0.64}{
\begin{tabular}{|l|l|l|l|l|l|l|l|l|l|l|l|l|l|}
\hline
ROI   & \begin{tabular}[c]{@{}l@{}}rh G \\ subcallosal\end{tabular} & \begin{tabular}[c]{@{}l@{}}rh S orbital\\  med.olfact\end{tabular} & \begin{tabular}[c]{@{}l@{}}lh G \\ precuneus\end{tabular} & \begin{tabular}[c]{@{}l@{}}lh G temp \\ sup.Lateral\end{tabular} & \begin{tabular}[c]{@{}l@{}}lh S \\ suborbital\end{tabular} & \begin{tabular}[c]{@{}l@{}}rh S cingul.\\ Marginalis\end{tabular} & \begin{tabular}[c]{@{}l@{}}lh G and S \\ cingul.Mid.Post\end{tabular} & \begin{tabular}[c]{@{}l@{}}rh S oc\\ .temp lat\end{tabular} & \begin{tabular}[c]{@{}l@{}}lh G and S\\  paracentral\end{tabular} & \begin{tabular}[c]{@{}l@{}}lh G and S\\  subcentral\end{tabular} & \begin{tabular}[c]{@{}l@{}}lh S collat \\ transv post\end{tabular} & \begin{tabular}[c]{@{}l@{}}rh G Ins lg \\ and S cent ins\end{tabular} & rh G rectus \\ \hline
Thick & 0                                                           & 0                                                                  & 0                                                         & 0                                                                & 0                                                          & 0                                                                 & 1                                                                     & 1                                                           & 0                                                                 & 0                                                                & 0                                                                  & 0                                                                     & 0           \\ \hline
Vol   & 1                                                           & 1                                                                  & 0                                                         & 0                                                                & 0                                                          & 0                                                                 & 0                                                                     & 0                                                           & 1                                                                 & 1                                                                & 1                                                                  & 1                                                                     & 1           \\ \hline
Area  & 1                                                           & 1                                                                  & 0                                                         & 0                                                                & 0                                                          & 0                                                                 & 0                                                                     & 0                                                           & 0                                                                 & 0                                                                & 0                                                                  & 0                                                                     & 0           \\ \hline
Curv  & 0                                                           & 0                                                                  & 1                                                         & 1                                                                & 1                                                          & 1                                                                 & 0                                                                     & 0                                                           & 0                                                                 & 0                                                                & 0                                                                  & 0                                                                     & 0           \\ \hline
\end{tabular}
}
\vspace{-0.1cm}
\caption{
\small{
Upper table: Quantitative results of different methods. Lower table: Features related to change of SRS selected by SIS using a2009s atlas in FreeSurfer, 1 (0) represents selected (unselected)}}
\end{table*}
 \vspace{-0.3cm}
\section{Experiments}
\vspace{-0.2cm}
\subsection{Participants and measures}
\label{sec:subjects}
Nineteen children (13 males, 6 females, mean age = 5.87 years, s.d. = 1.09 years) with ASD participated in 16 weeks of PRT treatment. 
IQ was measured using the Wechsler Abbreviated Scale of Intelligence (WASI). All participants were highly functioning (IQ  $\geq$ 70, Mean IQ  =  104.5, SD = 16.7) regarding full-scale IQ. All participants met the diagnostic criteria for ASD determined by the results of the Autism Diagnostic Observation Schedule (ADOS) \cite{lord1989autism}. The regression targets($y$) are the differences between pre and post-treatment scores, including ADOS calibrated severity score, ADOS social affect total score, ADOS restricted and repetitive behavior total score and social responsiveness scale (SRS) total score \cite{constantino2012social}. 
\vspace{-0.3cm}
\subsection{Imaging acquisition and processing}
\label{freesurfer}
Each child underwent pre-treatment and post-treatment scans on a Siemens MAGNETOM 3T Tim Trio scanner. A structural MRI image series was acquired with a 12-channel head coil and a high-resolution T1-weighted MPRAGE sequence with the following imaging parameters: 176 slices, TR = 2530 ms, TE = 3.31ms, flip angle = 7 \degree, slice thickness = 1.0 mm, voxel size = $1 \times 1 \times 1 mm^3$, matrix = $256 \times 256$. 
\par
The structural MRI was processed with FreeSurfer \cite{fischl2012freesurfer} using the "recon-all" command with 31 stages of processing, and the key steps include: (a) motion correction and conform, (b) non-uniform intensity normalization, (c) skull strip and removal of neck, (d) registration, (e) white matter segmentation, (f) cortical parcellation, (g) cortical parcellation statistics. We used freesurfer\_stats2table\_bash commands from FreeSurfer official website to extract structural statistics. Four features of each region of interest (ROI) in the Destrieux atlas \cite{destrieux2010automatic} were extracted, including volume, cortical thickness, surface area and mean curvature, resulting in a total number of 592 features (4 features $\times$ 148 cortical ROIs). 
\vspace{-0.3cm}
\subsection{Predictive modeling and method comparison}
We predicted treatment outcomes (changes in ASD severity scores) from changes of structural information (post-treatment structural features minus pre-treatment features) extracted from FreeSurfer, and we included phenotype information such as age, gender and pre-treatment IQ in our model as confounding factors. We performed leave-one-out cross-validation (LOOCV) and measured cross-correlation and root mean squared error (RMSE) between prediction and ground-truth outcomes. We chose LOOCV instead of correlation analysis to validate the predictability of selected features and reduce the false discovery rate. 
\par
The analysis was conducted in R with default parameters unless stated. We used ``gaussian" family in package ``SIS'', and set ``penalty'' as ``MCP" and set ``tune" as ``bic". Besides SIS, we applied other linear regression models as comparison, including: (a) elastic-net regression with nested LOOCV to select parameter $\lambda$ (100 default values) and $\alpha$ (ranging from 0 to 1 with a step size of 0.1) with package ``glmnet", (b) support vector regression (SVR)\cite{basak2007support} with package ``e1701", and (c) partial least squares regression (PLSR) \cite{geladi1986partial} with nested LOOCV to select the number of components (ranging from 1 to maximum number of components) with package ``pls". 
\vspace{-0.3cm}
\section{Results}
\begin{figure}[!htb]
    \centering
    \begin{minipage}{.23\textwidth}
        \centering
        \includegraphics[width=1\linewidth]{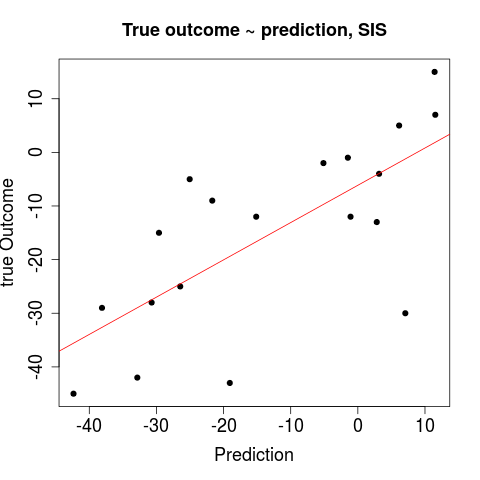}
        
    \end{minipage}%
    \begin{minipage}{0.23\textwidth}
        \centering
        \includegraphics[width=1\linewidth]{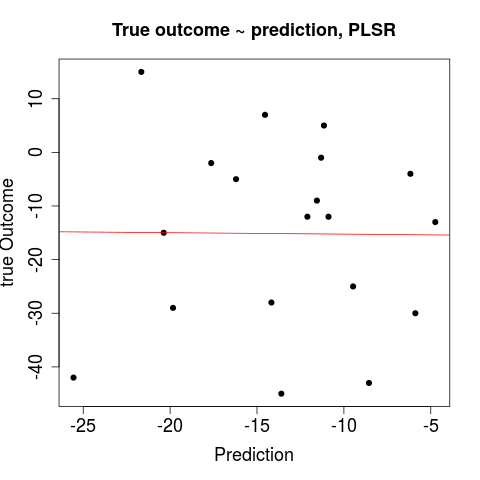}
       
    \end{minipage}
    \begin{minipage}{.23\textwidth}
        \centering
        \includegraphics[width=1\linewidth]{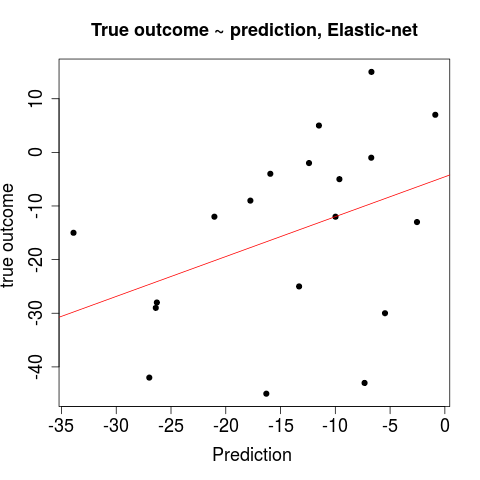}
        
    \end{minipage}%
    \begin{minipage}{.23\textwidth}
        \centering
        \includegraphics[width=1\linewidth]{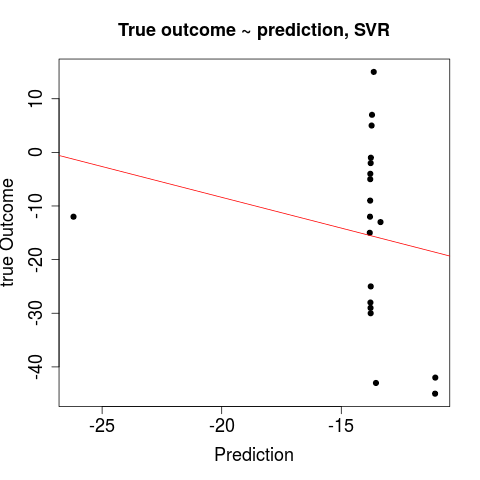}
        
    \end{minipage}%
    \caption{
    \footnotesize{Prediction of change of SRS score from different methods}
    }
    \label{SRS}
\end{figure}

\vspace{-0.2cm}
Results of different methods on predicting changes of SRS scores are shown in Fig.\ref{SRS}. Compared to other methods, results from SIS method lie nearest to the ideal line $prediction =  measurement$. In the high-dimensional case, it's easy to select features that have a high correlation with responses of training data but lack predictive ability on test data. Sometimes the model even produces a negative correlation as shown in Fig. \ref{SRS} with PLSR and SVR, because the maximum spurious correlation grows with dimensionality, while SIS produces predictive results in the cross-validation.

\begin{figure}[!htb]
    \centering
    \begin{minipage}{.25\textwidth}
        \centering
        \includegraphics[width=1\linewidth]{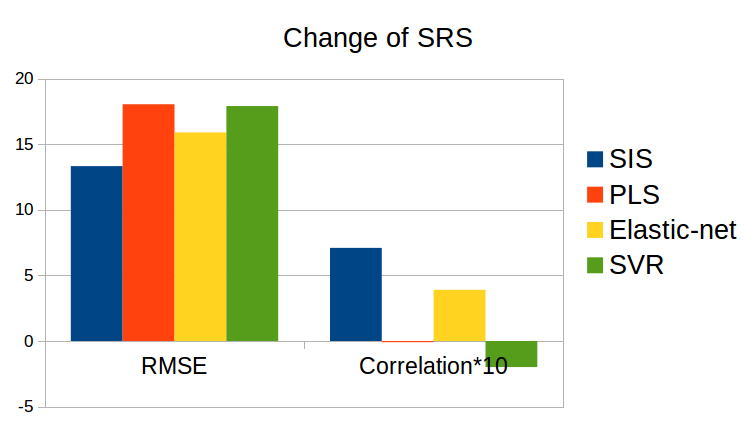}
        
    \end{minipage}%
    \begin{minipage}{0.25\textwidth}
        \centering
        \includegraphics[width=1\linewidth]{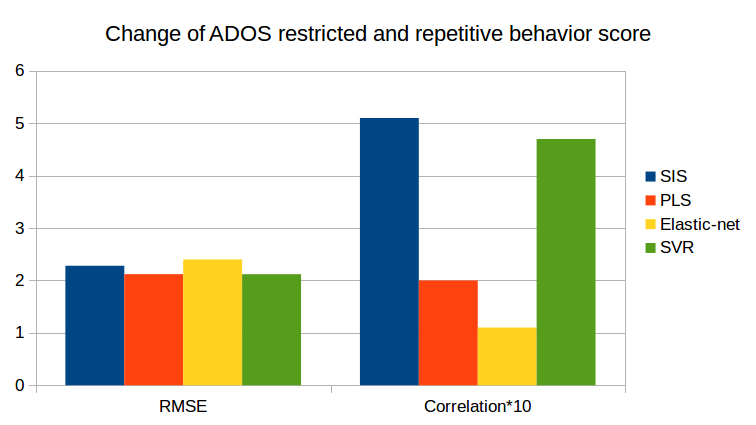}
       
    \end{minipage}
    \begin{minipage}{.25\textwidth}
        \centering
        \includegraphics[width=1\linewidth]{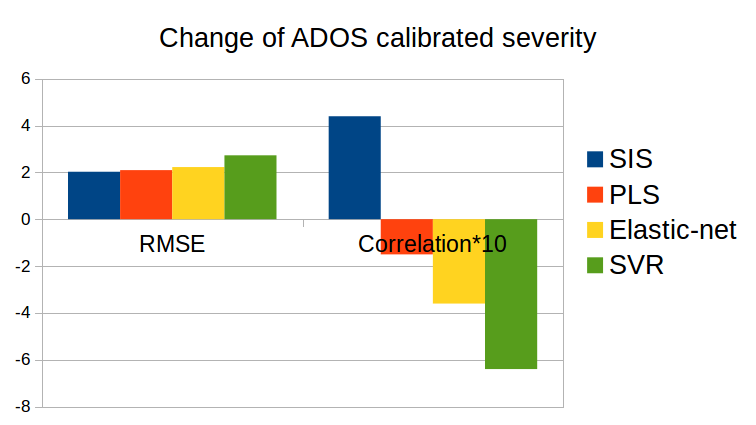}
        
    \end{minipage}%
    \begin{minipage}{.25\textwidth}
        \centering
        \includegraphics[width=1\linewidth]{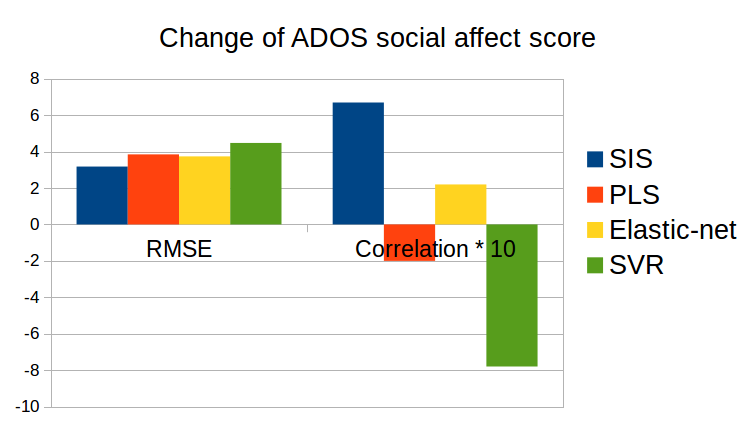}
        
    \end{minipage}%
    \caption{
    \footnotesize{Performance of different methods on different tasks. RMSE and Correlation are reported. Correlations are multiplied by 10 for display purpose.}
    }
    \label{comparison}
\end{figure} 
\vspace{-0.4cm}
\par
Performance of different methods on various tasks are summarized in Fig.\ref{comparison} and the upper part of table 1.  RMSE and correlation between prediction and ground-truth outcomes are reported. Corrleation is multiplied by 10 for display purpose. Compared with other methods, SIS produces the highest correlation in all tasks, and it produces a correlation of above 0.4 in ADOS calibrated severity task, while all other methods generate negative correlations because the wrong features are selected. SIS generates the lowest RMSE in the prediction of changes in SRS, ADOS calibrated severity and ADOS social affect score. For results on ADOS restricted and repetitive behavior, four methods produce very similar RMSE, while SIS generates the highest correlation. 
\par
Features selected by SIS on the SRS task are reported in the lower part of table 1. Our findings match previous literature: group differences between ASD and control have been found in anterior subcallosal gyrus \cite{paakki2010alterations}, precuneus \cite{martineau2010atypical}, cingulate gyrus, paracentral lobule, superior frontal gyrus and paracentral gyrus\cite{pereira2018differences}. The selected features are not only correlated with treatment outcomes, but also are predictive for treatment outcomes. 
\vspace{-0.5cm}
\section{Discussion and conclusion}
\vspace{-0.5cm}
The high dimensionality is a huge difficulty in many medical image analysis problems. We introduce SIS for general linear models in the ultra-high dimensional case, and validate its superior performance over traditional methods on different tasks. SIS selects structural changes in the brain that are predictive for treatment outcomes, which is useful for understanding the effect of behavioral treatments.
\subsubsection*{Acknowledgement}
This research
was funded by the National Institutes of Health (NINDS-R01NS035193).
\bibliographystyle{IEEEbib}
\begin{frame}{}
    \footnotesize{
\bibliography{refs}
}
\end{frame}

\end{document}